\documentclass{elsart}
\usepackage{amssymb}
\usepackage{graphicx}
\begin{document}
\begin{frontmatter}

\title{Nucleon form factors in the canonically quantized Skyrme model}
\author[itpa]{A. Acus}
\author[itpa]{E. Norvai\v{s}as}
\author[hels,hip]{D.O. Riska}
\address[itpa]{Institute of Theoretical Physics and Astronomy,
Vilnius, 2600 Lithuania}
\address[hels]{Department of Physics, University of Helsinki, 00014
Finland}
\address[hip]{Helsinki Institute of Physics}

\begin{abstract}
The explicit expressions for the
electric, magnetic, axial and induced pseudoscalar form 
factors of the nucleons are derived
in the {\it ab initio} quantized Skyrme model. The canonical 
quantization procedure ensures 
the existence of stable soliton solutions
with good quantum numbers. 
The form factors are derived for representations of arbitrary
dimension of the $SU(2)$ group.
After fixing the two parameters of the model, 
$f_\pi$ and $e$, by the empirical mass and 
electric mean square radius of the proton, the calculated 
electric and magnetic form factors 
are fairly close to the empirical ones, whereas the 
the axial and induced pseudoscalar form factors fall 
off too slowly with momentum transfer.
\end{abstract}

\end{frontmatter}

\section{Introduction}
\label{intro}
Quantum chromodynamics (QCD) describes the nucleon as
a composite system with many internal degrees of 
freedom. In the nonperturbative region, which encompasses
hadron structure and intermediate range observables, 
the large $N_C$ limit of the theory, which partly 
allows treatment in closed form, has proven to be of phenomenological
utility. This limit of QCD may be realized either in 
terms of the
constituent quark model, or, as was first suggested, in the form of 
effective meson field theory, in which the baryons appear as topologically
stable soliton solutions \cite{Jenkins}.

The generic chiral topological soliton model with topologically
stable solutions, which
represent
baryons is that of T.H.R. Skyrme \cite{Sky1}. The first
comprehensive 
phenomenological application of the model to nucleon structure
was the semiclassical calculation of the 
static properties of the nucleon in ref. \cite{Adk}. 
In terms of agreement with the static baryon observables the results
fell short by typically $\sim$ 30 $\%$, which could be viewed as an
expected flaw
of the large $N_C$ limit.  That approach to the model did however
have the more principal imperfection in that its lack of stable 
semiclassical solutions with good quantum numbers. 
The quantization was therefore
realised by means of rigid body 
rotation of the classical skyrmion solutions. 
A strictly canonical quantization of Skyrme model was 
derived later in ref.\cite{Fujii} and shown to yield   
stable quantized skyrmion solutions in refs.\cite{Nor1,Acus97,Acus98}.
In addition the Skyrme model was generalized to representations 
of arbitrary dimension of the SU(2) group. 
It was shown that - in fact an obvious consistency check - the
classical skyrmion solutions are independent of the dimension of
the representation, but that in contrast the canonically quantized
Skyrme model gives results for baryon observables, which 
are representation dependent.

An interesting consequence of the canonical {\em ab initio\/} quantization 
of the Skyrme model is the natural appearance of a finite 
effective pion mass even for the chirally symmetric 
Lagrangian. While the finite pion mass is conventionally
introduced by adding an explicitly chiral 
symmetry breaking pion mass term to the Lagrangian density of the 
model \cite{Nappi},
the canonical quantization procedure by itself gives rise to a finite
pion mass. This realizes Skyrme's original
conjecture that "This (chiral) symmetry is, however, destroyed by the 
boundary condition ($U(\infty )=1$), and we believe that the mass (of pion) 
may arise as a self consistent quantal effect \cite{Sky2}". 

To derive the explicit expressions for electric, magnetic, axial and 
pseudoscalar form factors of the nucleon we employ the expressions
for the Noether currents derived in ref.\cite{Acus98}. 
Because of the appearance of a finite "effective"
pion mass the asymptotic behavior of the  
chiral angle $F(r)$ has the required exponential falloff,
which ensures finite radii and physical forms for
the energy (mass) density. The expressions for the current
operators 
are valid for representations of arbitrary 
dimension of SU(2). Numerical results
are given for the 
representations with $j = 1/2; 1; 3/2$ and also for the
reducible representation $j = 1\oplus 1/2\oplus 1/2\oplus 0$. 
The different representations of the quantized Skyrme model
may be interpreted as different phenomenological models. The best 
agreement with experimental data on the form factors obtain
with the 
reducible SU(2) representation, which in fact is the SU(3) group octet 
$(1,1)$ restricted to the SU(2).

This paper is divided into 5 sections. In Section~\ref{sec2} the canonically
quantized skyrmion is 
reviewed. In Section~\ref{sec3} we derive the electroweak form factors of 
the nucleon.  
Section~\ref{sec4} contains the numerical results and 
Section~\ref{sec5} a concluding discussion.

\vspace{1cm}

\section{Canonically quantized skyrmion}
\label{sec2}
The chirally symmetric Lagrangian density that defines the
Skyrme model may be written in the form \cite{Adk}:
\begin{equation}
{\mathcal{L}}[U({{\bf r}},t)]=-\frac{f_\pi ^2}{4}\mathrm{Tr}\{R_\mu R^\mu 
\}+{\frac1{32e^2}\mathrm{Tr}}\{[R_\mu ,R_\nu ]^2\},
\label{f1}
\end{equation}
where $R_\mu$ is the "right handed" chiral current
$R_\mu=(\partial _\mu U)U^{\dagger }$. The 
unitary field $U({{\bf r}},t)$ may, in a general  
reducible representation of the SU(2) group, be expressed as a 
direct sum of Wigner's D matrices:
\begin{equation}
U({{\bf r}}, t)=\sum_k \oplus D^{j_k}[\vec\alpha({{\bf r}},t)].
\label{f0}
\end{equation}
Here the vector
$\vec\alpha$ represents a triplet of Euler angles $\alpha_1({\bf r},t)$,
$\alpha_2({\bf r},t)$, $\alpha_3({\bf r},t)$.  

Quantization of  
the skyrmion field $U$ is brought about by means of rotation
by collective coordinates that separate the variables, which
depend on time and spatial coordinates:
\begin{equation}
U({\bf r},{\bf q}(t))=
A\left( {\bf q}(t)\right) U_0({\bf r})A^{\dagger}\left( {\bf q}(t)\right).
\label{f2} 
\end{equation}
Here the matrix $U_0$ is the generalization of the
classical hedgehog ansatz to a general 
reducible representation \cite{Acus98}. The 
collective coordinates 
 ${\bf q}(t)$ (the Euler angles) are dynamical variables  
that satisfy the commutation relations
$[\dot q^a,\,q^b]\neq 0$. The energy of the canonically
quantized skyrmion, which represents a baryon with spin-isospin 
$\ell $, which corresponds to the Lagrangian density~(\ref{f1})  
in an arbitrary reducible representation has the form:
\begin{equation}
E(j,\ell ,F)=M(F)+\Delta M_j(F)+\frac{\ell (\ell +1)}{2a(F)},
\label{f3} 
\end{equation}
where $M(F)$ represents the classical skyrmion mass:

\begin{equation}
M(F)=2\pi \frac{f_\pi}{e}\int \d\tilde{r}\tilde{r}^2\biggl( F^{\prime
2}+\frac{\sin ^2F}{\tilde{r}^2}\Bigl( 2+2F^{\prime 2}
+\frac{\sin ^2F}{\tilde{r}^2}\Bigr) \biggr).
\label{f4} 
\end{equation}
The dimensionless variable 
$\tilde{r}=ef_\pi r$ has been employed here.
Above $a(F)$ represents the moment of inertia of the 
skyrmion:
\begin{equation}
a(F)=\frac{8\pi }3\frac{1}{e^3f_\pi }\int 
\d\tilde{r}\tilde{r}^2\sin ^2F\Bigl( 1
+F^{\prime 2}+\frac{\sin
^2F}{\tilde{r}^2}\Bigr),
\label{f5} 
\end{equation}
and $\Delta M_j(F)$ is a (negative) mass term, which appears
in the canonically quantized version of the model:
\begin{eqnarray}
\Delta M_j(F)&=&\frac{-2\pi }
{15e^3f_\pi {a}^2(F)}\int \d\tilde{r}\tilde{r}^2\sin ^2F 
\Bigl(15+4d_2\sin ^2F(1-F^{\prime 2}) \label{f6} \\
&&\phantom{
\frac{-2\pi }{5e^3f_\pi {a}^2(F)}\int \d\tilde{r}\tilde{r}^2\sin ^2F 
\Bigl(15}
+2d_3\frac{\sin ^2F}{\tilde r^2}
+2d_1F^{\prime 2}\Bigr)\nonumber.
\end{eqnarray}
The coefficients $d_{i}$ in these expressions are given as 
\begin{eqnarray}
N&=&\frac23\sum_{k}j_{k}(j_{k}+1)(2j_{k}+1).  \label{F24a}\\
d_{1} &=&\frac{1}{N}\sum_{k}j_{k}(j_{k}+1)(2j_{k}+1)\bigl(
8j_{k}(j_{k}+1)-1\bigr) ,  \label{F24b} \\
d_{2} &=&\frac{1}{N}\sum_{k}j_{k}(j_{k}+1)(2j_{k}+1)(2j_{k}-1)(2j_{k}+3),
\label{F24c} \\
d_{3} &=&\frac{1}{N}\sum_{k}j_{k}(j_{k}+1)(2j_{k}+1)\bigl(
2j_{k}(j_{k}+1)+1\bigr) .  \label{F24d}
\end{eqnarray}
Minimization of the mass expression in Eq.~(\ref{f4}) for $M(F)$, 
gives the classical solution for the chiral angle 
$F(r)$, which behaves as $1/\tilde r^2$ at large distances. 
In the semiclassical 
case, the quantum mass correction $\Delta M_j(F)$ drops out,
and variation of the expression (\ref{f3}) yields no
stable solution \cite{Braaten}. Such  
a semiclassical skyrmion
was considered in ref.~\cite{Adk} as a "rotating"
rigid-body skyrmion with fixed $F(r)$. The canonical 
quantization procedure 
in terms the collective coordinates approach leads to the 
expanded energy expression ~(\ref{f3}), 
variation of which yields a (self-consistent) integro-differential 
equation with the 
boundary conditions $F(0)=\pi $ and $F(\infty )=0$.
In contrast to the semiclassical case, the asymptotic behavior
of 
$F(\tilde r)$ at large $\tilde r$ falls off exponentially as:
\begin{equation}
F(\tilde r)=k\left( \frac{\tilde m_\pi }{\tilde r}+\frac 1{\tilde 
r^2}\right) \exp(-\tilde m_\pi \tilde r),
\label{f12} 
\end{equation}
with
\begin{equation}
\tilde m_\pi ^2=-\frac{1}{3e^2f_\pi^2 a(F)}\left( 8\Delta 
{M_j}(F)+\frac{2\ell (\ell +1)+3}{ a(F)}\right).
\label{f11}
\end{equation}
The integrals~(\ref{f4}), (\ref{f5}), (\ref{f6}) are convergent,
and therefore ensure the 
stability of the  quantum skyrmion only if $\tilde m_\pi ^2>0$. The positive 
quantity $m_\pi =ef_\pi \tilde m_\pi $ admits an obvious 
interpretation as an effective 
pion mass. The appearance of this effective pion
mass conforms with Skyrme's original conjecture
concerning the origin of the pion mass.

\section{Form factors}
\label{sec3}
The electroweak form factors of the 
semiclassically quantized SU(2) skyrmion were studied
systematically in ref.\cite{Braaten1}. An extension of this work to the SU(3) 
was made in ref. \cite{Meissner}. Analogous studies of the
electroweak form factors in the related 
SU(3) chiral Quark-Soliton Model has been made in ref.\cite{Praszal}.

The explicit expressions for the Noether current density
operators of the canonically quantized
Skyrme model were derived in ref.\cite{Acus98}. The isoscalar part of the 
nucleon electromagnetic current operator is 
proportional to the topological baryon current operator
and therefore depends on the Lagrangian density only through
the chiral angle. 
The isovector component of the vector current of the nucleon 
current is proportional to vector Noether current
of the Lagrangian density~\cite{Acus98}. 
Linear combinations of the isoscalar and
isovector charge densities yield the expressions for
the proton and 
the neutron charge densities:
\begin{eqnarray}
\rho _{p}(r) &=&-\frac{1}{4\pi ^{2}r^{2}}F^{\prime }(r)\sin ^{2}F(r) 
\nonumber \\ 
&&\phantom{-}+\frac{1}{3a(F)}\sin ^{2}F(r)\biggl( f_{\pi }^{2}+\frac{1}{e^{2}}\Bigl( 
F^{\prime 2}(r)+\frac{\sin ^{2}F(r)}{r^{2}}\Bigr) \biggr),
\label{fa1} \\
\rho _{n}(r) &=&-\frac{1}{4\pi ^{2}r^{2}}F^{\prime }(r)\sin ^{2}F(r)
\nonumber \\
&&\phantom{-}-\frac{1}{3a(F)}\sin ^{2}F(r)\biggl( f_{\pi }^{2}+\frac{1}{e^{2}}\Bigl( 
F^{\prime 2}(r)+\frac{\sin ^{2}F(r)}{r^{2}}\Bigr) \biggr).
\label{fa2}
\end{eqnarray}
respectively.
The Fourier transforms of these charge densities, which are spherically 
symmetric scalar functions,
give the electric 
form factors of proton and the neutron in the Breit frame as:
\begin{equation}
G_{E}^{p}(Q^2)=
4\pi \int \d r r^{2}\rho _{p}(r)j_{0}(qr),
\label{fa3}
\end{equation}
\begin{equation}
G_{E}^{n}(Q^2)=
4\pi \int \d r r^{2}\rho _{n}(r)j_{0}(qr). 
\label{fa4}
\end{equation}
Here $j_{n}(qr)$ is the spherical Bessel function of n-th order
and Q is the \hbox{4-momentum} transfer to the nucleon ($Q^2=-{\bf q}^2$).

The isoscalar and isovector magnetic form factors for the nucleon may be 
expressed as
\begin{eqnarray}
G_{M}^{S}(Q^2)&=&\frac{-2m}{\pi a(F)q}\int \d r rF^{\prime }(r)
\sin ^{2}F(r)j_{1}(qr),
\label{fa5}
\\[3pt]
G_{M}^{V}(Q^2)&=&\frac{16\pi m}{3q}\int \d r r \biggl( f_{\pi 
}^{2}+\frac{1}{e^{2}}\Bigl( F^{\prime 2}(r)+\frac{\sin 
^{2}F(r)}{r^{2}}\nonumber \\
&&\phantom{\frac{16\pi m}{3q}\int \d r r \biggl( }
-\frac{2d_{2}-15}{4\cdot5a^{2}(F)}\sin ^{2}F(r)\Bigr) 
\biggr)\sin ^{2}F(r) j_{1}(qr).
\label{fa6}
\end{eqnarray}
Recombination into proton and neutron form factors yields
\begin{eqnarray}
G_{M}^{p}(Q^2)&=&\half\Bigl( G_{M}^{S}(Q^2)+G_{M}^{V}(Q^2)\Bigr),
\label{fa7} 
\\
G_{M}^{n}(Q^2)&=&\half\Bigl( G_{M}^{S}(Q^2)-G_{M}^{V}(Q^2)\Bigr).
\label{fa8} 
\end{eqnarray}
The magnetic form factors at zero-momentum transfer give the magnetic 
moments of nucleons as
\begin{equation}
G_{M}^{p}(0) =\mu _{p},\qquad
G_{M}^{n}(0) =\mu _{n},
\label{fa9}
\end{equation}
in units of nuclear magnetons.

The standard 
definition of the matrix element of the axial vector current
of the nucleon is
\begin{eqnarray}
\left\langle N^{\prime }({\bf p_2})\left| A_{\mu }^{i}(0)\right|
N({\bf p_1})\right\rangle &=&\overline{u}({\bf p_2})\tau ^{i}\biggl( 
\gamma _{\mu}\gamma _{5}G_{A}(Q^2)\label{fa10} \\
&&\phantom{\overline{u}({\bf p_2})\tau ^{i}\Bigl(\gamma _{\mu}\gamma _{5}}
+q_{\mu }\gamma _{5}\frac{G_{P}(Q^2)}
{2m}\biggr) u({\bf p_1}),\nonumber
\end{eqnarray}
where $G_A(Q^2)$ and $G_P(Q^2)$ are the axial vector and
induced pseudoscalar form factors respectively and 
${\bf q}={\bf p_2}-{\bf p_1}$

In the non-relativistic limit the axial current operator takes the form
\begin{eqnarray}
\left\langle N^{\prime }({\bf p_{2}})\left| A_{b}^{a}(0)\right|
N({\bf p_{1}})\right\rangle  &=&
\left\langle N^{\prime }\left| \tau ^{a}\sigma _{b^{\prime }}\right|
N\right\rangle
\biggl((-1)^{b}
\delta _{b,-b^{\prime }} G_{A}(Q^2)\label{fc4}
\\
&&
\phantom{\left\langle N^{\prime }\left| \tau ^{a}\sigma _{b^{\prime }}\right|
N\right\rangle\biggl((-1)}
-\frac{q^{2}}{4m^2}
\hat{q}_{b}\hat{q}_{b^{\prime }}G_{P}(Q^2)\biggr)
\nonumber
\end{eqnarray}

Here it is convenient to employ the circular coordinates system 
for spin and isospin. The momentum 
transfer 
${\bf q}=q\hat {\bf q}$ is then:
\begin{equation}
\hat{q}_{a}=\frac{2\sqrt{\pi }}{\sqrt{3}}Y_{1,a}(\vartheta ,\varphi ).
\label{fc3} 
\end{equation}
The induced pseudoscalar form factor now takes the form:
\begin{eqnarray}
G_{P}(Q^2) &=&-\frac{3\sqrt{2\cdot5}m^2}{\sqrt{\pi }q^{2}}
\int {\rm d}\vartheta {\rm d}\varphi 
\sin \vartheta \left\langle 
p\left|A_{0}^{1}(0)\right| n\right\rangle Y_{2,0}(\vartheta ,\varphi )
\nonumber \\
&=&-\frac{16\pi m^2}{3q^2}\int r^{2}\d r \bigg( f_{\pi }^{2}\Bigl( 
2F^{\prime }-\frac{\sin 2F}{r}\Bigr)-\frac{1}{e^{2}}\Bigl( F^{\prime 
2}\frac{\sin 2F}{r}\nonumber
\\
&&\phantom{-\frac{16\pi m^2}{3q^2}\int r^{2}\d r \bigg(}
-4F^{\prime }\frac{\sin^{2}F}{r^{2}} 
+\frac{\sin ^{2}F\sin 
2F}{r^{3}} 
\label{fc9} \\
&&
\phantom{-\frac{16\pi m^2}{3q^2}\int r^{2}\d r \bigg(}
+\frac{\sin ^{2}F\sin 2F}{4a^{2}(F)r}+F^{\prime }\frac{\sin 
^{2}F}{4a^{2}(F)}\Bigr) \biggr) j_{2}(qr).
\nonumber
\end{eqnarray}
The axial form factor is
\begin{eqnarray}
G_{A}(Q^2) &=&
\frac{1}{\sqrt{2\pi }}\int 
\d \vartheta \d \varphi  \sin \vartheta 
\left\langle p\left| A_{0}^{1}(0)\right|
n\right\rangle \Bigl( Y_{0,0}(\vartheta ,\varphi )-\frac{\sqrt{5}}{
2}Y_{2,0}(\vartheta ,\varphi )\Bigr)
\nonumber \\
&=&-\frac{8\pi }{9}\int r^{2}\d r  \biggl( f_{\pi 
}^{2}\Bigl( F^{\prime }+\frac{\sin 2F}{r}\Bigr) +\frac{1}{e^{2}}\Bigl( 
F^{\prime2}\frac{\sin 2F}{r}+2F^{\prime }\frac{\sin ^{2}F}{r^{2}}
\nonumber \\  
&&
\phantom{-\frac{8\pi }{9}\int r^{2}\d r  \biggl(}
+\frac{\sin ^{2}F\sin 2F}{r^{3}} -\frac{5\sin ^{2}F\sin 
2F}{4a^{2}(F)r}\label{fa11} \\
&&
\phantom{-\frac{8\pi }{9}\int r^{2}\d r  \biggl(}
-F^{\prime }\frac{\sin^{2}F}{8a^{2}(F)}\Bigr) \biggr)
j_{0}(qr) 
+\frac{q^2}{12m^2}G_{P}(Q^2).\nonumber
\end{eqnarray}
The expression (\ref{fa11}) equals that for the  
axial form factor given in ref.\cite{Nyman}, with exception
for the quantum corrections 
$\sim 1/a^{2}(F)$ which appear in the canonical quantization
procedure.

The axial current operator contains terms of
fourth order in the components of ${\bf r}$~\cite{Acus98}, and 
consequently its 
Fourier transform involves terms of fourth order in 
$ {\bf q} $. To avoid a redefinition of the
axial current (\ref{fa10}), we reduce 
it to $Y_{4,a}(\vartheta ,\phi )$, and terms of second and zero order
in the components of ${\bf q}$. 

The electromagnetic mean square radii of nucleons is determined by means of 
the expression:
\begin{equation}
\left\langle r^2 \right\rangle = -\frac{6}{G(0)}
\frac\d {\d q^2}G(-q^2)
\label{fb11}
\end{equation}
The effect of 
Lorentz boosts for these form factors may be taken into
account by means of the rescalings~\cite{Ji}:
\begin{eqnarray}
{}_{rel}G_{E}^{p,n}(Q^{2})&=&
G_{E}^{p,n}\left(\frac{Q^{2}}{1+Q^{2}/4m^{2}}\right), 
\label{fa12}
\\
{}_{rel}G_{M}^{p,n}(Q^{2})&=&
\frac{1}{1+Q^{2}/4m^{2}}G_{M}^{p,n}\left(\frac{Q^{2}}{1+Q^{2}/4m^{2}}\right),
\label{fa13} 
\\
{}_{rel}G_{A}(Q^{2})&=&
\frac{1}{\sqrt{1+Q^{2}/4m^{2}}}G_{A}\left( 
\frac{Q^{2}}{1+Q^{2}/4m^{2}}\right),
\label{fa14} 
\\
{}_{rel}G_{P}(Q^{2})&=&
\frac{1}{\sqrt{(1+Q^{2}/4m^{2})^{3}}}G_{P}\left( 
\frac{Q^{2}}{1+Q^{2}/4m^{2}}\right).
\label{fa15} 
\end{eqnarray}
These boost corrections are numerically significant for large
values of momentum transfer.

\section{Numerical results}
\label{sec4}
\begin{table}
\begin{center}
\caption{The predicted static baryon observables for different
representations with fixed empirical values for the proton radius
$\left\langle r^2\right\rangle_E^p=0.735$~fm$^2$ 
and nucleon mass $939$~MeV.
The experimental data on the nucleon mean square radii are from 
ref.\cite{Grash}. }
\label{table}
\begin{center}
\begin{tabular}[h]{|c|c|c|c|c|c|c|}
\hline \hline
$\mathbf{j}$ & \textbf{Classical~\cite{Nappi}}&
$\mathbf{1/2}$ & $\scriptscriptstyle \mathbf{1\oplus \frac 12 \oplus \frac
12}$& $\mathbf{1}$ & $\mathbf{3/2}$ &\textbf{Expt.} \\ \hline
$m$&Input\footnotemark&Input&Input &Input&Input&$939$~MeV \\ \hline
$\left\langle r^2\right\rangle^p_E$&$\infty$&Input&Input &Input&Input&$0.735$~fm$^2$ \\ \hline
$f_\pi$& 64.5 & 64.8 & 60.3& 59.4 & 57.5 & $93$~MeV\\ \hline
$e$& 5.45 & 4.76 & 4.31& 4.19 & 3.86 & \\ \hline
$\left\langle r^2\right\rangle^n_E$&$\infty$&-0.368&-0.269&-0.249&-0.210&-0.114~fm$^2$\\ \hline
$\left\langle r^2\right\rangle^p_M$&$\infty$&0.618&0.594&0.587&0.575&0.719~fm$^2$\\ \hline
$\left\langle r^2\right\rangle^n_M$&$\infty$&0.687&0.609&0.594&0.567&0.637~fm$^2$\\ \hline
$\mu_p$& 1.87 & 1.96 & 2.32& 2.39 & 2.54 & 2.79\\ \hline
$\mu_n$& -1.31 & -1.37 & -1.73& -1.81 & -1.99 & -1.91\\ \hline
$g_A$& 0.61 & 0.73& 0.84& 0.87 & 0.95 & 1.26\\ \hline
$m_\pi$& 0 & 110& 171& 191 & 246 & 138~MeV \\ \hline \hline
\end{tabular}
\end{center}
\end{center}
\end{table}
\footnotetext{Ref.\cite{Nappi} used the $\Delta$ 
resonance mass $1232$~MeV as an input parameter.}

The nucleon form factors have been calculated numerically
in the representations of the SU(2) group with $j = 
1/2, 1, 3/2$  and in the reducible 
representation $1 \oplus 1/2 \oplus 1/2 \oplus 0$. The two parameters of the
Lagrangian density, $f_\pi$ and $e$, have been determined here
so that the empirical mass of the proton (938 MeV) and its
electric mean square radius (0.735 fm$^2$) are reproduced for
each value of $j$ (Table~\ref{table}). The chiral angle $F(r)$ for each 
one of these 
representations has been determined by self consistent  
numerical variation of the energy expression (\ref{f3}). 
This procedure yields four pairs of model parameters $f_\pi$ and $e$,
all of which are close to the values in~\cite{Acus98}.

The value of the axial coupling constant $g_A$, which 
is far too small in the semiclassical version of
the Skyrme model remains below 1 in all the representations
considered. The reason for this systematic underestimate is
the absence of a quark contribution to the helicity of the
nucleon as 
explained by a sum rule argument in ref.\cite{Mari}.
The "effective" pion mass $m_{\pi}$ describes the behavior at infinity of 
the chiral angle $F(R)$ and the asymptotic 
falloff $e^{-2m_{\pi}r}$ of nucleon 
mass density.

\begin{figure}
\begin{center}
\includegraphics*[scale=1.0]{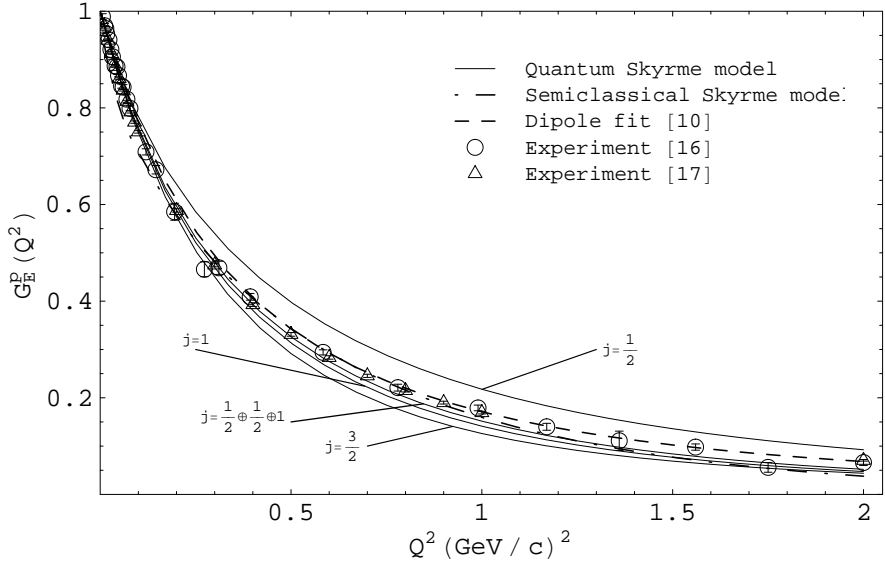}
\end{center}
\caption{
Proton electric form factor $G_E^p(Q^2)$ with relativistic corrections.}
\label{pe}
\end{figure}

The calculated electric form factor of the proton 
as obtained with the boost corrections
(\ref{fa12}) are plotted in Fig.~\ref{pe}. 
In this case the form factor that is calculated in 
the reducible representation
comes closest to the dipole fit to the empirical data.

\begin{figure}
\begin{center}
\includegraphics*[scale=1.0]{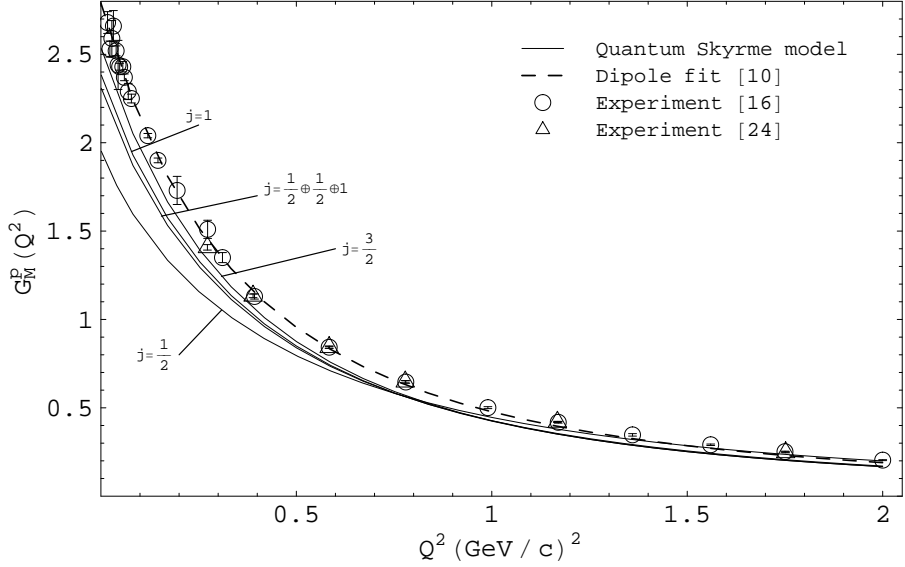}
\end{center}
\caption{
Proton magnetic form factor $G_M^p(Q^2)$ with relativistic corrections.}
\label{pm}
\end{figure}

The corresponding magnetic form factors of the proton, again
including the boost correction (\ref{fa13}), are plotted
in Fig.~\ref{pm}. In this case all the calculated form factors
have a realistic falloff with momentum transfer at low
values of momentum transfer, although the absolute
predictions for the magnetic moment of the proton fall
short by some $\sim$ 10-20\%.
In the semiclassical case the magnetic form factor 
is not well defined \cite{Adk}.

\begin{figure}
\begin{center}
\includegraphics*[scale=1.0]{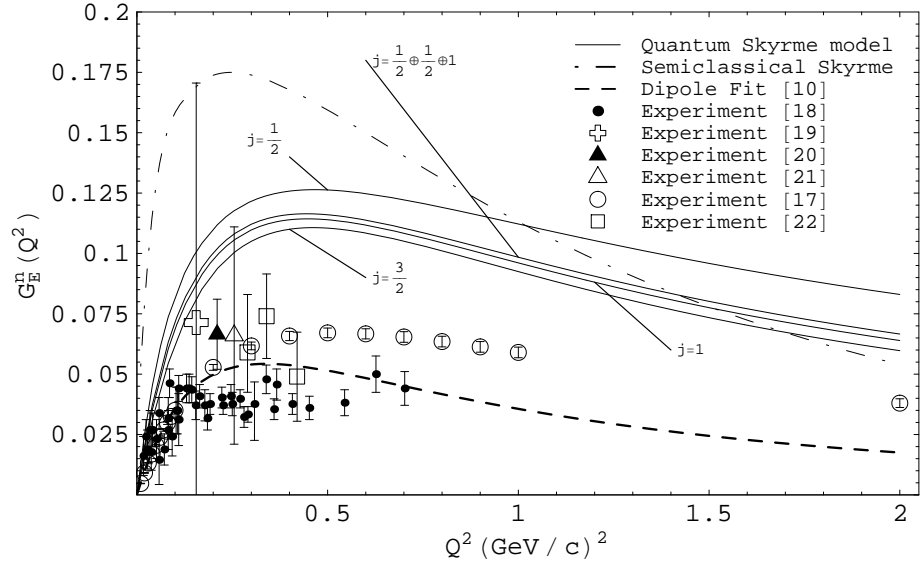}
\end{center}
\caption{
Neutron electric form factor $G_E^n(Q^2)$ with relativistic corrections.}
\label{ne}
\end{figure}
In Fig.~\ref{ne} the calculated electric form factors of neutron are
shown. The results again include the boost
correction (\ref{fa12}). 
The experimental data in this case have too wide uncertainty margins
for model discrimination. The new 
experimental results obtained by polarized electron 
scattering 
\cite{Pass,Ost} indicates this form factor to much larger than
what earlier data have suggested, and thus closer to the
present calculated values, even though these are still much
larger than the empirical results at intermediate values
of momentum transfer.

\begin{figure}
\begin{center}
\includegraphics*[scale=1.0]{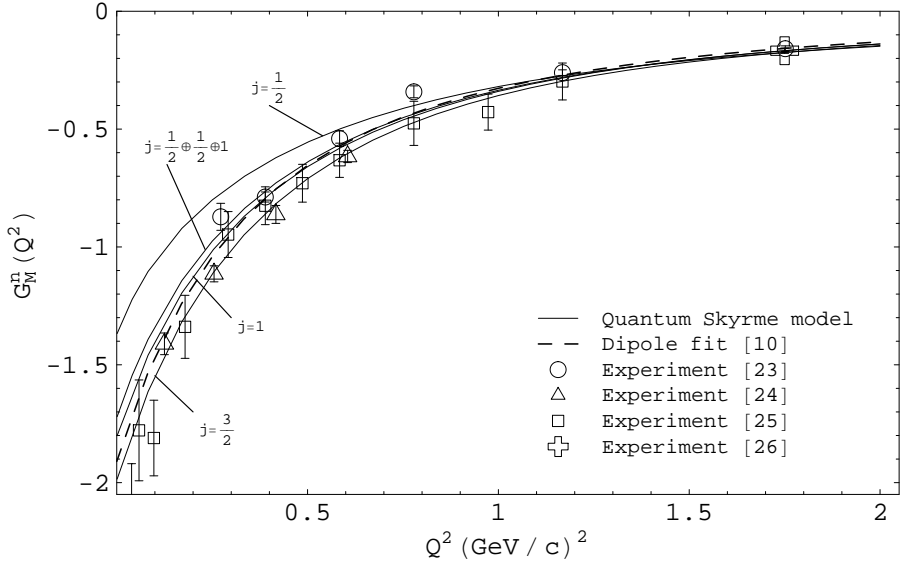}
\end{center}
\caption{
Neutron magnetic form factor $G_M^n(Q^2)$ with relativistic corrections.}
\label{nm}
\end{figure}
In Fig.~\ref{nm} we plot the magnetic form factors of neutron as
obtained with the boost
correction (\ref{fa13}). 
In terms of agreement with the empirical form factor
values only the results for the fundamental representation
in which $j=1/2$ is found to be wanting.
This form factor is also ill defined in the semiclassical
case.

\begin{figure}
\begin{center}
\includegraphics*[scale=1.0]{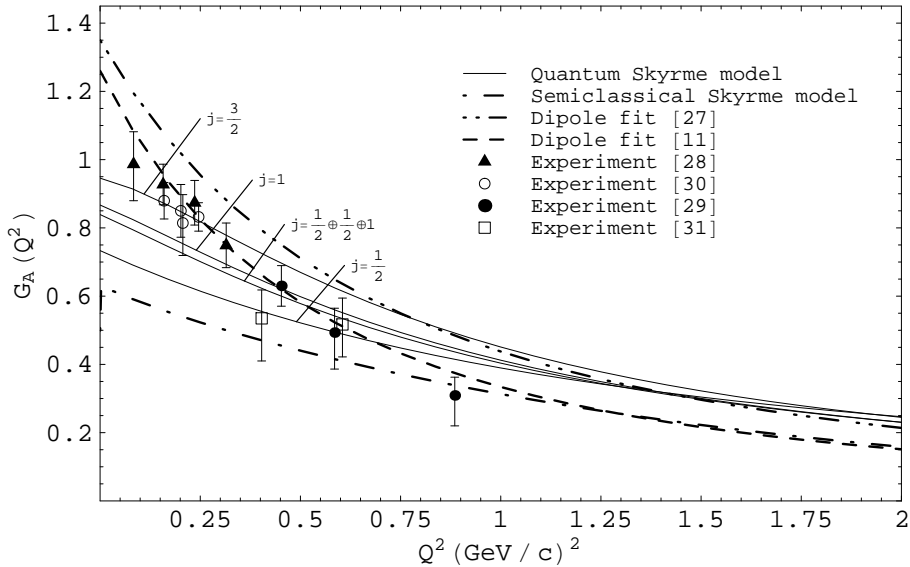}
\end{center}
\caption{
Nucleon axial form factor $G_A(Q^2)$ with relativistic corrections.}
\label{ga}
\end{figure}
In Fig.~\ref{ga} we plot the axial form factor of nucleon with 
the boost 
correction (\ref{fa14}). 
The empirical values for the axial form factor have a 
dipole-like 
behavior. The Skyrme model form factors tend to underestimate
the falloff rate with momentum considerably, although it is
possible to find parameter values that bring the axial
coupling constant close to the empirical value in the case
of the quantum skyrmion.

\begin{figure}
\begin{center}
\includegraphics*[scale=1.0]{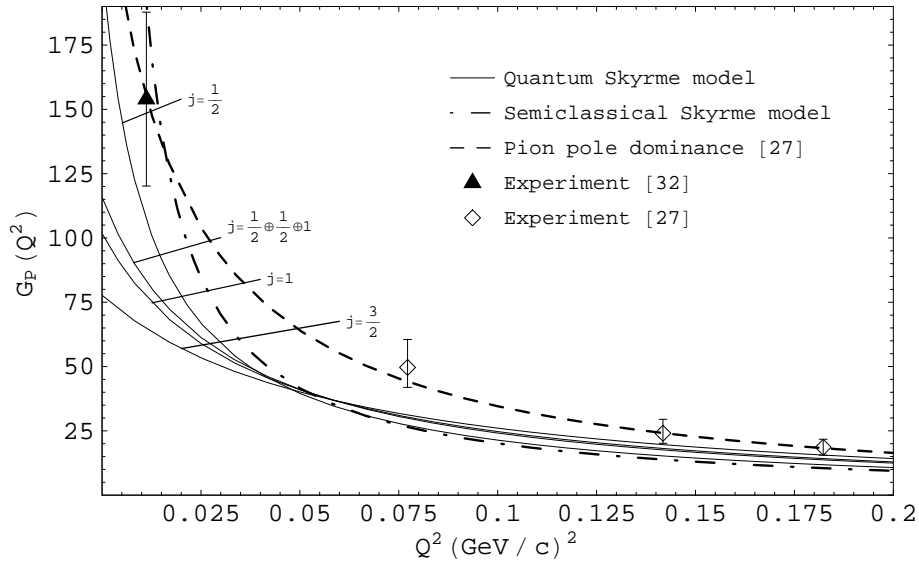}
\end{center}
\caption{
Nucleon pseudoscalar form factor $G_P(Q^2)$ with relativistic
corrections.}
\label{gpa}
\end{figure}
In Fig.~\ref{gpa} we plot the pseudoscalar form factor of nucleon with 
the boost correction. This correction represents only about a 
1\% correction at $Q^2$ = 0.2 
(GeV/c)$^2$ For this form factor the calculated values fall below the 
uncertainty margin of the experimental values, with exception of case of the 
semiclassical result, which is too large at small values of
momentum transfer and too small at large values.

Mathematica 
(Wolfram Research inc.) has been extensively used  both for
symbolic and numerical calculations \cite{wolf}.

\section{Discussion}
\label{sec5}
The nucleon form factors are well defined in the Skyrme model if the
chiral angle asymptotically falls faster than by the semiclassical rate
$1/r^2$. The desired exponential fall of has to be brought about by a
finite pion mass term, which implies breaking of chiral symmetry. While
the pion mass term may be introduced at the classical level through an
explicit chiral symmetry breaking term in the Lagrangian density, we
have previously shown that such breaking of chiral symmetry also arises,
without additional mass parameters, in the canonical {\em ab initio\/}
quantization of the Skyrme model \cite{Acus98}. As shown here, this
ensures well defined nucleon form factors, which - at least
in the case of the electromagnetic form factors - do have
phenomenologically adequate momentum dependence. It has also been noted
elsewhere and in another context, that quantum corrections may generate
a finite pion mass \cite{Torn}.

The present work develops the phenomenological application of the
original Skyrme model to representations of arbitrary dimension of the
$SU(2)$ group, and by imposing consistent canonical quantization. This
of course in no way exhaust the phenomenological freedom of the
Skyrme model with only pion fields: the possibility for generalization
of the Lagrangian to terms of higher order in the derivatives remains
largely unexplored. Expanded versions of the topological soliton
models, which besides the pion fields, also 
contain vector meson fields have
additional mass scales and thus the parameter freedom, which makes it
possible to achieve closer agreement with experiment \cite{Meissner}.

\begin{ack} 
Research supported in part by the Academy of Finland through
contract 44903.
\end{ack}


\begin{thebibliography}{99}

\bibitem{Jenkins} E. Jenkins, Ann. Rev. Nucl. Sci. 48 (1998) 81.

\bibitem{Sky1} T. H. R. Skyrme, Proc. Roy. Soc. A 260 (1961) 127.

\bibitem{Adk}  G.S. Adkins, C.R. Nappi, and E. Witten, Nucl. Phys. 
B 228 (1983) 552.

\bibitem{Fujii} K. Fujii, A. Kobushkin, K. Sato and N. Toyota, Phys. 
Rev. D 35 (1987) 1896.

\bibitem{Nor1}
E. Norvai\v sas and D.O. Riska, Physica Scripta. 50 (1994) 634.
\bibitem{Acus97}  A. Acus, E. Norvai\v sas and D.O. Riska, Nucl. Phys. 
A 614 (1997) 361.

\bibitem{Acus98}  A. Acus, E. Norvai\v sas and D.O. Riska, Phys. Rev. 
C 57 (1998) 2597.

\bibitem{Nappi} G.S. Adkins and C.R. Nappi, Nucl. Phys. 233 (1984) 109.

\bibitem{Sky2} T.H.R. Skyrme, Nucl. Phys. 31 (1962) 556.

\bibitem{Braaten} E. Braaten and J.P. Ralston, Phys. Rev. D 31 (1985) 598.

\bibitem{Braaten1} E. Braaten, Sze-Man Tse, Ch. Willcox, Phys. Rev. 
D 34 (1986) 1482.

\bibitem{Meissner}  U.-G. Meissner, N. Kaiser and W. Weise , Nucl. 
Phys. A 466 (1987) 685.

\bibitem{Praszal} M. Praszalowicz, T. Watabe, K. Goeke, Nucl. Phys. 
A 647 (1999) 49.

\bibitem{Nyman}  E.M. Nyman and D.O. Riska, Rep. Prog. Phys. 53 (1990) 1137.

\bibitem{Ji} X. Ji, Phys. Lett. B 254 (1991) 456.

\bibitem{Hoehler} G. H\"ohler {\it et al.}, Nucl. Phys. B 114 (1976) 505.

\bibitem{Grash} A.F. Grashin, I.B. Lukasevich, Yad. Fiz. 62 (1999) 1632.

\bibitem{Platch} S. Platchkov {\it et al.}, Nucl. Phys. A 510 (1990) 740.

\bibitem{Jones} C.E. Jones {\it et al.}, Phys. Rev. C 44 (1991) R571.

\bibitem{Pass} I. Passchier {\it et al.}, Phys. Rev. Lett. 82 (1999) 4988.

\bibitem{Eden} T. Eden {\it et al.}, Phys. Rev. C 50 (1994) R1749.

\bibitem{Ost} M. Ostrick {\it et al.}, Phys. Rev. Lett. 83 (1999) 276.

\bibitem{Bruins} E.E.W. Bruins {\it et al.}, Phys. Rev. Lett. 75 (1995) 21. 

\bibitem{Hanson} K.M. Hanson {\it et al.}, Phys. Rev. D 8 (1973) 753.

\bibitem{Hughes} E.B. Hughes {\it et al.}, Phys. Rev. 139, No 2B, (1965) 
B458.

\bibitem{Lung} A. Lang {\it et al.}, Phys. Rev. Lett. 70 (1993) 718.

\bibitem{Choi} S. Choi {\it et al.}, Phys. Rev. Lett. 71 (1993) 3927.

\bibitem{Del} A.Del Guerra {\it et al.}, Nucl. Phys. B 99 (1975) 253.

\bibitem{DelG} A.Del Guerra {\it et al.}, Nucl. Phys. B 107 (1976) 65.

\bibitem{Amaldi} E. Amaldi {\it et al.}, Nuovo Cimento A 65 (1970) 
377; Phys. Lett. B 41 (1972) 216. 

\bibitem{Brauel} P. Brauel {\it et al.}, Phys. Lett. B 45 (1973) 389; 
Phys. Lett. B 50 (1974) 507.

\bibitem{Esaulov} A.S. Esaulov, A.M. Pilipenko, and Yu.I. Titov, Nucl. 
Phys. B 136 (1978) 511.

\bibitem{Torn} N.A. T\"ornqvist, Phys. Lett. B 426 (1998) 105.
\bibitem{Mari} M. Kirchbach and D. O. Riska, Nuovo Cim, {\bf A104}
(1991) 1837

\bibitem{wolf} S. Wolfram, The Mathematica Book, 4th ed. (Wolfram 
Media/Cambridge University Press, 1999)

\end{thebibliography}
\end{document}